\documentclass[12pt]{article}
\usepackage{bbm, url}
\renewcommand{\theequation}{\thesection.\arabic{equation}}
\makeatletter
\@addtoreset{equation}{section}
\makeatother

\usepackage{amssymb,mathrsfs,multirow,color}

\def\Eqn#1{Eq.\ (\ref{#1})}
\def\Eqs#1#2{Eqs.\ (\ref{#1}) and (\ref{#2})}
\def\3Eqs#1#2#3{Eqs.\ (\ref{#1}), (\ref{#2}) and (\ref{#3})}
\def\sec#1{\S\ref{#1}}

\def\next{\nonumber\\}
\def\unitmat{\mathbbm 1}
\def\comm#1,#2/{\Big[ #1, #2 \Big]}

\title{\bf Reduction formulas for symmetric products of spin matrices}

\author{\bf Palash B. Pal\\
Saha Institute of Nuclear Physics\\ 
1/AF Bidhan-Nagar, Calcutta 700064, India}

\date{}

\begin{document}

\maketitle

\begin{abstract}

We show that, for SU(2) generators of arbitrary dimension $D$, there
exist identities that express the completely symmetric product of $D$
matrices in terms of completely symmetric products of fewer number of
matrices.  We also indicate why such identities are important in
characterizing electromagnetic interactions of particles.

\end{abstract}

\section{Introduction}
The purpose of this article is to present several identities involving
spin matrices, or equivalently SU(2) generators \cite{su2}.  The
identities are independent of any basis used for writing the explicit
forms of the generators, and can be used to write products of a
certain number of spin matrices by using products of smaller number of
such matrices.

The SU(2) generators, or the spin matrices, satisfy the commutation
relation~\cite{hbar}
\begin{eqnarray}
\comm S_i, S_j/ = i \sum_k \varepsilon_{ijk} S_k 
\label{comm}
\end{eqnarray}
where the summation \cite{sum} over $k$ runs from 1 to 3 since there
are three generators, and $\varepsilon_{ijk}$ is the completely
antisymmetric symbol with
\begin{eqnarray}
\varepsilon_{123} = +1 \,.
\end{eqnarray}

\Eqn{comm} shows that an antisymmetric product of two spin matrices
yields terms with only one spin matrix.  From this, it is easy to show
that if any expression containing spin matrices is antisymmetric in
two indices, the expression can be written by using smaller number of
spin matrices.  For example, consider the string $S_i S_j S_k - S_k
S_j S_i$, which is obviously antisymmetric under the interchange of
the indices $i$ and $k$.  Also notice that
\begin{eqnarray}
S_i S_j S_k - S_k S_j S_i 
&=& \comm S_i,S_j/ S_k + S_j \comm S_i, S_k/ + \comm S_j, S_k/ S_i \next
&=& i \sum_l \left( \varepsilon_{ijl} S_l S_k + \varepsilon_{ikl} S_j
S_l +  \varepsilon_{jkl} S_l S_i \right) \,,
\label{ijk}
\end{eqnarray}
using \Eqn{comm} to write the last step.  This example shows that the
product of three spin matrices, antisymmetrized in two indices, can be
expressed as terms which contain products of only two spin matrices.
It is easy to generalize this result to products of arbitrary number
of spin matrices which are antisymmetric with respect to the
interchange of at least one pair of indices.

It therefore remains to be seen whether the same kind of reduction is
possible for other kinds of products of spin matrices, which are not
antisymmetric under the interchange of any pair of indices.  It
suffices to examine the possibility for completely symmetric products,
because any product can be written in terms of symmetric and
antisymmetric combinations.

We will show that reduction identities exist among completely
symmetric combinations of spin matrices.  However, unlike
\Eqs{comm}{ijk} which are obeyed by all representations of the spin
matrices, the identities that follow depend on the dimensionality of
the representations.  In order to present these identities, we will
use curly brackets to denote symmetric combination of any number of
spin matrices.  Thus, for example, with two spin matrices, the
symmetric product is denoted by
\begin{eqnarray}
\{ S_{i} S_{j} \} &\equiv& S_i S_j + S_j S_i \,, 
\label{symm2}
\end{eqnarray}
which is just the anticommutator.  With three matrices, we define
\begin{eqnarray}
\{ S_i S_j S_k \} &\equiv& S_i S_j S_k + S_j S_k S_i + S_k S_i S_j +
S_i S_k S_j + S_k S_j S_i + S_j S_i S_k \next
&=& \{ S_i S_j \} S_k + \{ S_j S_k \} S_i + \{ S_k S_i \} S_j \,.
\label{symm3}
\end{eqnarray}
The generalization is obvious.  For example, the symmetric combination
with four spin matrices is given by
\begin{eqnarray}
\{ S_i S_j S_k S_l \} &\equiv& \{ S_i S_j S_k \} S_l + \{ S_j S_k
S_l \} S_i +  \{ S_k S_l S_i \} S_j +  \{ S_l S_i S_j \} S_k \,.  
\label{symm4}
\end{eqnarray}
In addition, we will also use the Kronecker delta, as well as its
generalizations involving more than two indices, such as
\begin{eqnarray}
\delta_{ijkl} \equiv \delta_{ij} \delta_{kl} +  \delta_{ik}
  \delta_{jl} + \delta_{il} \delta_{jk} \,.
\end{eqnarray}
For the generalized Kronecker delta with $2n$ indices, each term
contains a product of $n$ Kronecker deltas, and the number of such
terms is
\begin{eqnarray}
T_n \equiv 1 \cdot 3 \cdot 5 \cdots (2n-1) = {(2n)! \over 2^n \, n!}
\;. 
\end{eqnarray}
%

\section{Examples with various representations}\label{s:ex}
We start with the examples of identities of symmetric combinations of
spin matrices.  As commented above, these identities depend on the
dimension of the representation, so we discuss various low-dimensional
representations one by one.

\subsection{2-dimensional representation}
The generators for the 2-dimensional representation are $\frac12
\sigma_i$, where $\sigma_i$ ($i=1,2,3$) are the Pauli matrices.  It is
well-known that the square of each Pauli matrix is the unit matrix,
and different Pauli matrices anticommute.  Both kinds of information
can be encoded by writing
\begin{eqnarray}
2 \{ S_i S_j \} - \delta_{ij} \unitmat = 0 \,.
\label{2did}
\end{eqnarray}
This identity reduces the symmetric combination of two spin matrices
to a multiple of the unit matrix, which does not contain any spin
matrix at all.  Note that the symmetric combinations of larger number
of spin matrices can all be built up from the symmetric combination of
two spin matrices, as has been shown in \Eqs{symm3}{symm4}.  Thus,
\Eqn{2did} guarantees that symmetric combination of any number of
2-dimensional spin matrices can be reduced to smaller number of
matrices.  This general comment is true for all representations ---
once we obtain an identity involving a certain number of spin
matrices, it guarantees that there are such identities with larger
number of spin matrices.  This comment will not be repeated for higher
representations.

\subsection{3-dimensional representation}
For 3-dimensional representation, we cannot find any identity for the
symmetric combination of two spin matrices, or the anticommutator.
With three spin matrices, however, we find the identity
\begin{eqnarray}
\{ S_i S_j S_k \} - 2 \Big( S_i \delta_{jk} + S_j \delta_{ki} + S_k
\delta_{ij} \Big) = 0 \,.
\label{3did}
\end{eqnarray}
If $i,j,k$ denote the same index in \Eqn{3did}, the identity reduces
to 
\begin{eqnarray}
S_i^3 = S_i \,,
\end{eqnarray}
which is the obvious characteristic equation of the 3-dimensional spin
matrices since the eigenvalues have to be $0$ and $\pm1$.  However,
\Eqn{3did} is much more general than this characteristic equation.
For example, if two of the three indices appearing in \Eqn{3did} are
equal and the other is different, we obtain the relation
\begin{eqnarray}
S_i^2 S_j + S_i S_j S_i + S_j S_i^2 = 2 S_j, \qquad (i\neq j) \,.
\end{eqnarray}
And finally, if all three indices are different, \Eqn{3did} reduces to
the equation
\begin{eqnarray}
S_1 S_2 S_3 + S_2 S_3 S_1 + S_3 S_1 S_2 + S_1 S_3 S_2 + S_3 S_2 S_1 +
S_2 S_1 S_3 = 0 \,.
\end{eqnarray}
%

\subsection{4-dimensional representation}
In this case, the lowest order identity involves symmetric combination
of four spin matrices, and the identity is
\begin{eqnarray}
2 \{ S_i S_j S_k S_l \} - 10 \Big( \{ S_i S_j \} \delta_{kl} +
  \mbox{(5 more similar terms)} \Big) + 9 \delta_{ijkl} \unitmat = 0 \,.
\label{4did}
\end{eqnarray}
For $i=j=k=l$, this equation reduces to
\begin{eqnarray}
16 S_i^4 - 40 S_i^2 + 9 \cdot \unitmat = 0 \,,
\end{eqnarray}
which is easily seen to be the characteristic equation of the spin
matrices whose eigenvalues are $\pm\frac12$ and $\pm\frac32$.  In
addition, \Eqn{4did} implies many other equations.  For example, with
$i\neq j$, we obtain the following identities:
\begin{eqnarray}
8 \Big( S_i^2 S_j^2 + S_j^2 S_i^2 + S_i S_j^2 S_i + S_j S_i^2 S_j +
S_i S_j S_i S_j + S_j S_i S_j S_i \Big) \next
- 20 \Big( S_i^2 + S_j^2 \Big)
+ 9 \cdot \unitmat &=& 0 \,, \\ 
2 \Big( S_i S_j^3 + S_j S_i S_j^2 + S_j^2 S_i S_j + S_j^3 S_i \Big) -
5 \Big( S_i S_j + S_j S_i \Big) &=& 0 \,.
\end{eqnarray}
There will also be similar identities that involve three different
indices $i,j,k$, of the form
\begin{eqnarray}
2 \Big( 
S_i^2S_jS_k + S_i^2S_kS_j + S_iS_jS_iS_k + S_iS_jS_kS_i +
S_iS_kS_iS_j + S_iS_kS_jS_i \qquad \next*
+ S_jS_i^2S_k + S_jS_iS_kS_i + 
S_jS_kS_i^2 + S_kS_i^2S_j + S_kS_iS_jS_i + S_kS_jS_i^2 
  \Big) \qquad \next*
- 5 \Big( S_j S_k + S_k S_j \Big) = 0 \,.
\end{eqnarray}
%

\subsection{5-dimensional representation}
In this case, the simplest identity involves products of five spin
matrices:
\begin{eqnarray}
\{ S_i S_j S_k S_l S_m \} - 10 \Big( \{ S_i S_j S_k \} \delta_{lm}
+ \mbox{(9 more similar terms)} \Big) && \next* 
+ 32 \Big( S_i \delta_{jklm} + \mbox{(4 more similar terms)} \Big) &=&
0 \,. 
\label{5did}
\end{eqnarray}
As with the previous cases, this identity contains the characteristic
equation: 
\begin{eqnarray}
S_i^5 - 5 S_i^3 + 4 S_i = 0 \,,
\end{eqnarray}
but it also entails many other identities for cases where all indices
are not equal.

\section{General remarks}
It should be noticed that all identities given above are independent
of the basis in which the spin matrices are written.  If we change the
basis, the matrix $S_i$ changes to
\begin{eqnarray}
S'_i = US_i U^\dagger
\end{eqnarray}
for some unitary matrix $U$.  It is easily seen that such
transformations have no effect on the identities.  More generally, any
transformation of the form
\begin{eqnarray}
S_i' = M S_i M^{-1} \,,
\end{eqnarray}
with any non-singular matrix $M$, does not change the commutation
relation of \Eqn{comm} as well as the identities given in \sec{s:ex}.

In the examples given above, note that for $D$-dimensional
representation of the spin matrices, the simplest identity that we
obtain with the symmetric combinations contains a combination of order
$D$.  It is easy to see that it must be so.  If we put all indices to
be equal in such equations, we obtain an equation that the eigenvalues
of the spin matrices must satisfy.  Spin matrices in $D$-dimensional
representation have $D$ distinct eigenvalues, so the equation must be
of degree $D$.

It should be noted that the eigenvalues of a spin matrix of dimension
$D$, with even $D$, contains all eigenvalues of a spin matrix whose
dimension is also even, but smaller than $D$.  The same can be said
about odd $D$.  As a result, \Eqn{4did} will also be satisfied by
2-dimensional spin matrices, and \Eqn{5did} by the 3-dimensional spin
matrices.  In fact, starting from definitions like \Eqn{symm4} and
using \Eqn{2did}, one can easily show that \Eqn{4did} is satisfied.
Similarly, one can build one's way from \Eqn{3did} to \Eqn{5did}.

The numerical co-efficients appearing in various identities are simple
combinatorial factors.  These can be worked out from the
characteristic equations for the spin matrices.  Let us write the
characteristic equation in a form where the co-efficient of the term
with the highest power is unity:
\begin{eqnarray}
S_i^D + \sum_{p=1}^{\lfloor D/2 \rfloor} a_p S_i^{D-2p} = 0 \,,
\label{cheqn}
\end{eqnarray}
where $\lfloor D/2 \rfloor$ denotes the largest integer equal to or
less than $D/2$.  The generalized identity involving symmetric
combinations can then be written as
\begin{eqnarray}
\{ S_{i_1} S_{i_2} \cdots S_{i_D} \} + \sum_{p=1}^{\lfloor D/2 \rfloor} b_p
\Big( \{ S_{i_1} \cdots S_{i_{D-2p}} \} \delta_{i_{D-2p+1} \cdots
  i_D} + \mbox{similar terms} \Big) = 0 \,,
\label{genid}
\end{eqnarray}
where
\begin{eqnarray}
b_p = 2^p \, p! \; a_p \,.
\label{bp}
\end{eqnarray}
The number of ``similar terms'' in \Eqn{genid} is easily seen to be
${D \choose 2p}-1$, since there are ${D \choose 2p}$ ways of choosing
the indices in the symmetric product of $D-2p$ spin matrices, out of
which one has already been shown explicitly.

In fact, one can do better than \Eqn{bp}.  One can find the $b_p$'s in
closed form by putting in the analytical expressions for the $a_p$'s.
The latter can be found from the characteristic equations.  If the
largest eigenvalue $L$ equals $n$ where $n$ is a non-negative integer,
the characteristic equation will be of the form
\begin{eqnarray}
S_i (S_i^2 - 1^2) (S_i^2 - 2^2) \cdots (S_i^2 - n^2) = 0 \,.
\end{eqnarray}
On the other hand,  for half-integral spins, denoting the largest
eigenvalue by $L=n+\frac12$, the characteristic equation can be
written as
\begin{eqnarray}
\Big( S_i^2 - \frac{1^2}{2^2} \Big) \Big( S_i^2 - \frac{3^2}{2^2}
\Big) \cdots 
\Big( S_i^2 - \frac{(2n+1)^2}{2^2} \Big) = 0 \,.
\end{eqnarray}
Thus, for example, the co-efficient $a_1$ appearing in \Eqn{cheqn} is
easily seen to be
\begin{eqnarray}
a_1 = \cases { - \Sigma_2(n) & for
$L=n$\,, \cr  
- \Big( \Sigma_2(n) + \Sigma_1(n) + \frac14 \Sigma_0(n) \Big) \qquad
& for  $L=n+\frac12$\,,} 
\label{a1}
\end{eqnarray}
where we have used the symbol
\begin{eqnarray}
\Sigma_r(n) \equiv \sum_{q=0}^n q^r \,.
\end{eqnarray}

The co-efficient of the term with the smallest power of $S_i$ can also
be written down easily.  For odd $D$, i.e., for $L=n$, the last term
is $a_n S_i$, where
\begin{eqnarray}
a_n = (-1)^n (n!)^2 \,.
\end{eqnarray}
For even $D$, i.e., for $L=n+\frac12$, the last term is
\begin{eqnarray}
a_{n+1} = \left( {(2n+1)!! \over 2^{n+1}} \right)^2 = \left( {(2n+1)! 
  \over 2^{2n+1} \, n!} \right)^2 \,. 
\end{eqnarray}
However, as we move towards the middle of the series of terms that
appear in \Eqn{cheqn}, general expressions for the co-efficients look
more and more cumbersome, although they can all be written down in
closed form.  For example, for $L=n$, the co-efficient $a_2$ is given by
\begin{eqnarray}
a_2 = \sum_{q=0}^{n-1} \sum_{q'=q+1}^n q^2 q'^2 = 
\Big( \Sigma_2(n) \Big)^2 - \sum_{q=0}^n q^2 \Sigma_2(q) \,.
\end{eqnarray}
The summation will involve the sum of three or higher powers of
natural numbers.  All such sums can be carried out, as shown in the
Appendix.

\section{Utility}
Are these identities useful?  In other words, can they be used to
tackle some physics problem?  We can think of one such utility.

If we put a spin-$\frac12$ fermion like the electron in a magnetic
field $\vec B$, the electron experiences an interaction proportional
to the field strength.  The fact is summarized by saying that the
electron has a magnetic dipole moment.  The dipole moment is
proportional to spin, the only inherent property of a particle that
transforms like a vector, so that the interaction can be written as
some constant times $\vec S \cdot \vec B$.  One might ask, why doesn't
the electron have any quadrupole moment?  If it did, there would have
been interactions of the form
\begin{eqnarray}
H_{\rm int} = \mbox{(constant)} \times \sum_{i,j} \{S_iS_j\} {\partial
F_i \over \partial x_j} \,,
\label{quadru}
\end{eqnarray}
where $F_i$ could be either the electric field or the magnetic field.
It is clear why the symmetric product of the two spin vectors is used:
the antisymmetric combination could have been reduced to only one spin
matrix through \Eqn{comm}, and the resulting interaction would be none
other than the dipole interaction.  However, the symmetric product can
be reduced by \Eqn{2did}, so this interaction is not a quardrupole
interaction at all.  By the same argument, a spin-1 particle can have
a quadrupole moment in addition to a dipole moment, and therefore can
have an interaction term with the electromagnetic field of the form
shown in \Eqn{quadru}.  Such interactions were identified in a quantum
field theoretical investigation of the electromagnetic vertex of
spin-1 particles \cite{Nieves:1996ff}.  However, a spin-1 particle
cannot have any higher moment, because the symmetric product with
three spin matrices can be reduced to products of smaller number of
matrices, as shown in \Eqn{3did}.  Spin-$\frac32$ particles can also
have octupole moments, and so on.

It is true that the utility mentioned above depends not on the
explicit form of any identity but on the fact that such identities
exist.  It would be of interest to see whether the explicit
forms of various identities shown in \sec{s:ex} can be important in
some other physical context.

\bigskip

I thank Kumar S.\ Gupta for reading an earlier version of the
manuscript and commenting on it.

\appendix

\setcounter{equation}0
\renewcommand{\theequation}{A.\arabic{equation}}
\section*{Appendix: Finding the sums $\Sigma_r(n)$}
For $r=0$, we trivially obtain
\begin{eqnarray}
\Sigma_0(n) = n+1 \,.
\label{Sigma0}
\end{eqnarray}
In order to find $\Sigma_r(n)$ for an integer $r \geq 1$, we start
from the identity
\begin{eqnarray}
\sum_{q=0}^n q^{r+1} = \sum_{q=0}^n (q+1)^{r+1} - (n+1)^{r+1} \,.
\end{eqnarray}
Writing the binomial series for the power of $q+1$, we
obtain~\cite{GKP}
\begin{eqnarray}
\Sigma_r(n) = {1 \over r+1} \left[ (n+1)^{r+1} - \sum_{p=0}^{r-1} {r+1
  \choose p} \Sigma_p (n) \right] \,.
\end{eqnarray}
Starting from \Eqn{Sigma0}, one can now climb the ladder to the sums
of higher powers \cite{wolfram}: 
\begin{eqnarray}
\Sigma_1(n) &=& \frac12 n (n+1) \,, \\ 
\Sigma_2(n) &=& \frac16 n (n+1) (2n+1) \,, \\ 
\Sigma_3(n) &=& \frac14 n^2 (n+1)^2 \,, \\ 
\Sigma_4(n) &=& \frac1{30} n (n+1) (2n+1) (3n^2+3n-1)  \,, 
\end{eqnarray}
and so on.


\begin{thebibliography}{9}
\bibitem{su2} For an introduction to the group SU(2) and its
  representations, see textbooks on group theory.  Textbooks on
  quantum mechanics also discuss SU(2) representations.

\bibitem{hbar} Strictly speaking, the spin matrices are $\hbar$ times
  the SU(2) generators, and therefore a factor of $\hbar$ should
  appear on the right side of \Eqn{comm} if we are talking of the spin
  matrices.  We use the phrases ``SU(2) generators'' and ``spin
  matrices'' interchangeably, thus assuming a system of units in which
  $\hbar=1$. 

\bibitem{sum} Note that nowhere in this article we assume automatic
  summation over repeated indices.  All summations are indicated
  explicitly.

\bibitem{Nieves:1996ff} 
  J.~F.~Nieves and P.~B.~Pal,
  Phys.\ Rev.\ D {\bf 55}, 3118 (1997)
  [hep-ph/9611431].

\bibitem{GKP} This is a trivial generalization of one of the methods
  used for evaluating $\Sigma_2(n)$ given in R.~L. Graham, D.~E. Knuth
  and O. Patashnik, {\em Concrete Mathematics: A Foundation for
    Computer Science}, (Addison-Wesley; 2nd edition, 1994).

\bibitem{wolfram} Sums up to the 10th power appear on the internet
  site \url{ http:// mathworld . wolfram . com / PowerSum . html}

\end{thebibliography}
\end{document}